\documentclass[a4paper,12pt]{article}
\usepackage[left=2.5cm,bottom=3cm,right=2.5cm,top=3cm]{geometry} 

\usepackage{overpic,youngtab}
\usepackage{subfigure}
\usepackage[latin,english]{babel}
\usepackage{amsmath}
\usepackage{amssymb}
\usepackage{epstopdf}
\usepackage{graphics,psfrag,rotating}
\usepackage{mathabx}
\usepackage{graphicx}
\usepackage{dcolumn}
\usepackage{float}
\usepackage{pdflscape}
\usepackage{array}
\usepackage{booktabs}
\usepackage{amscd} 
\usepackage{mathtools}
\usepackage{fancybox}
\usepackage{tensor}
\usepackage{fix-cm}
\usepackage[colorlinks=true,citecolor=blue,,linktocpage=true,linkcolor=blue,urlcolor=black]{hyperref}
\usepackage{tikz} 
\usepackage{cleveref}
\usetikzlibrary{matrix} 
\usetikzlibrary{positioning} 
\usetikzlibrary{arrows}
\usepackage{titlesec}
\usepackage{abstract}
\usepackage{cancel}
\def\be{\begin{equation}}
\def\ee{\end{equation}}
\def\bea{\begin{eqnarray}}
\def\eea{\end{eqnarray}}

\def\a{\alpha}
\def\b{\beta}
\def\g{\gamma}
\def\d{\delta}
\def\m{\mu}
\def\n{\nu}
\def\t{\tau}

\def\l{\lambda}

\def\r{\rho}

\def\s{\sigma}
\def\e{\epsilon}
\def\bi{\begin{itemize}}
	\def\ei{\end{itemize}}

\begin{document}

		\vspace*{-1cm}
		\phantom{hep-ph/***} 
		{\flushleft
			{{FTUAM-xx-xx}}
			\hfill{{ IFT-UAM/CSIC-23-6}}}
		\vskip 1.5cm
		\begin{center}
		{\LARGE\bfseries  Variations on the Goroff-Sagnotti operator.}\\[3mm]
			\vskip .3cm
		
		\end{center}

		\vskip 0.5  cm
		\begin{center}
			{\large Enrique \'Alvarez, Jes\'us Anero and Eduardo Velasco-Aja.}
			\\
			\vskip .7cm
			{
				Departamento de F\'isica Te\'orica and Instituto de F\'{\i}sica Te\'orica, \\
				IFT-UAM/CSIC,\\
				Universidad Aut\'onoma de Madrid, Cantoblanco, 28049, Madrid, Spain\\
				\vskip .1cm

				\vskip .5cm
				
				\begin{minipage}[l]{.9\textwidth}
					\begin{center} 
							\textit{E-mail:} 
						\tt{enrique.alvarez@uam.es},
						\tt{jesusanero@gmail.com} and
						\tt{eduardo.velasco@uam.es}
					\end{center}
				\end{minipage}
			}
		\end{center}
	\thispagestyle{empty}
	
\begin{abstract}
	\noindent
The effect of modifying General Relativity with the addition of some higher dimensional operators, generalizations of the Goroff-Sagnotti operator, is discussed. 
We determine in particular, the general solution of the classical equations of motion, assuming it to be  spherically symmetric, not necessarily static. 
Even in the non-spherically symmetric case, we present a necessary condition for an algebraically generic spacetime to solve the corresponding equations of motion.	
Some examples of an application of said condition are explicitly worked out.
\end{abstract}

\newpage
\tableofcontents
	\thispagestyle{empty}
\flushbottom

\section{Prolegomena. }
In \cite{Alvarez2022} we have explored some consequences of the incorporation of the Goroff-Sagnotti counterterm to the Einstein-Hilbert action. Let us remind the reader that the Goroff-Sagnotti \cite{Goroff} counterterm is the first nontrivial renormalization of Einstein-Hilbert's lagrangian. It first appears at the two-loop order because  general relativity is one-loop finite, as shown in \cite{thooft}. 
To be specific,
\be
{\cal O}_{GS}\equiv W^{\m\n}\,_{\r\s}\,W^{\r\s}\,_{\g\d}\, W^{\g\d}\,_{\m\n},
\ee
where $W^{\m\n}\,_{\r\s}$ is Weyl's tensor\footnote{
Let us recall the definition of Weyl's tensor as
\bea
W_{\a\b\g\d}\equiv R_{\a\b\g\d}-{1\over n-2}\bigg\{g_{\a\g} R_{\b\d}-g_{\a\d} R_{\b\g}-g_{\b\g}R_{\d\a}+g_{\b\d} R_{\g\a}\bigg\}+{1\over (n-1)(n-2)}\,R\,\bigg\{g_{\a\g} g_{\d\b}-g_{\a\d} g_{\g\b}\bigg\}.\nonumber
\eea
}.
The structure of this operator is suggestive of a simple generalization. The purpose of the present letter is to study these more general operators, to wit
\be
 {\cal O}_{p}\equiv \text{tr}\,W^p\equiv W_{\m_1 \n_1 \r_1 \s_1}\ldots W_{\m_p\n_p \r_p \s_p} I^{\vec{\m}\vec{\n}\vec{\r}\vec{\s}},
\ee
where $I^{\vec{\m}\vec{\n}\vec{\r}\vec{\s}}$ is the product of $2p$ inverse metric tensors, for example
\be
I^{\vec{\m}\vec{\n}\vec{\r}\vec{\s}}\equiv g^{\m_1\m_2}\ldots g^{\m_{p-1}\m_p}g^{\n_1 \n_2} \ldots g^{\n_{p-1}\n_p}g^{\r_1\r_2}\ldots g^{\r_{p-1}\r_p} g^{\s_1\s_2}\ldots g^{\s_{p-1}\s_p},
\ee
although more complicated contractions are possible, they are all similar in character.

The main property of Weyl's tensor is that it is inert\footnote{Bear in mind the position of the indices in \cref{6}.} under the so-called Weyl rescalings (a gauge transformation) of the spacetime metric
\begin{align}
&g_{\m\n}\rightarrow \Omega(x)^2 g_{\m\n},\\
&W^\m\,_{\n\r\s}\rightarrow W^\m\,_{\n\r\s}.\label{6} 
\end{align}
\par
This means that  $S_p\equiv \int d(vol) {\cal O}_p$ (where $d(vol)\equiv \sqrt{|g|} d^n x$), is conformally invariant whenever $n=2p$.
\par
It would seem that, at least for compact spacetimes, the corresponding action $S_p$ should coincide with the integral of Branson's Q-curvature \cite{Branson}
\be
S_Q\equiv \int d(vol) Q,
\ee
whose variation is Fefferman and Graham's \cite{FG} obstruction tensor (a generalization of sorts of Bach's tensor \cite{Bach}  to arbitrary even dimension). But in fact, Alexakis' theorem \cite{Alexakis} proving Deser and Schwimmer's \cite{DS} conjecture guarantees that Branson's curvature is a linear combination of conformal invariants constructed out of Riemann tensor and its covariant derivatives plus total derivatives and Euler's density (the pfaffian).
\be
 Q= Weyl+ Pfaff+\nabla_\m j^\m.
 \ee
 Unfortunately, starting in $n=6$ \cite{Alvarez:2003, Bonora} there are other conformally invariant operators which are not Weyl polynomials. This means that starting on this dimension, our former conjecture ceases to hold.
 \par
 The operators ${\cal O}_p$ are, however, the only invariants not involving any derivative of Riemann's tensor.
 \par
 The main interest of those operators from our point of view is that they are natural generalizations of the Goroff-Sagnotti counterterm \cite{Goroff}. Some of the consequences of incorporating this operator in the classical action have been studied in a previous paper \cite{Alvarez2022}. The main purpose of this letter is to study generalizations of the type $p\geq 3$.
 \par
  Also, in the preceding work, we concentrated mainly on the spherically symmetric case; here we attempt to make more general statements. We shall examine in detail a particular equation (the trace equation) which is a necessary (and in many cases sufficient) condition for solving the equations of motion (EM).

\subsection{The operator ${\cal O}_2\equiv W^{\m\n}\,_{\a\b} W^{\a\b}\,_{\m\n}$. }
Let us now consider the four-dimensional conformal action
\be
S_{\tiny{Weyl^2}}\equiv \int d^4x \sqrt{|g|}W_{\m\n\r\s}W^{\m\n\r\s}.\label{CG} 
\ee  
In $n=4$ dimensions, using Lanczos' identity for the four-dimensional Weyl tensor \cite{Lanczos},
\be 
W_{\m\a\b\l}W_{\n}^{~\a\b\l} =\frac{1}{4}g_{\m\n} W^{\r\s}\,_{\a\b} W^{\a\b}\,_{\r\s},
\ee 
the EM are exactly \cite{Bach} Bach's tensor
\be 
{\d S_{\tiny{Weyl^2}}\over \d g^{\m\n}}=\sqrt{|g|}\Big\{\nabla^\l\nabla^\t W_{\l\m\t\n}+\frac{1}{2}W_{\l\m\t\n}R^{\l\t}\Big\}=\sqrt{|g|} B_{\m\n}=0,\ee
which is traceless;
\be B^\l_{~\l}=0.\label{12} \ee
Were we to consider this same action in arbitrary dimensions the EM are, 
\begin{equation}
	\frac{4 R^{\alpha \beta } W_{\mu \alpha \nu \beta }}{n-2}- \tfrac{1}{2} g_{\mu \nu } W_{\alpha \beta \gamma \epsilon } W^{\alpha \beta \gamma \epsilon }  + 2 W_{\mu }{}^{\alpha \beta \gamma } W_{\nu \alpha \beta \gamma } + 4 \nabla_{(\alpha }\nabla_{\beta) }W_{\mu }{}^{\alpha }{}_{\nu }{}^{\beta }=0,
\end{equation}
whose trace reads, 
\begin{equation}
	(2 -  \tfrac{1}{2} n) W_{\alpha \beta \gamma \epsilon } W^{\alpha \beta \gamma \epsilon }=0.
\end{equation}
We shall see later on the importance of this trace equation.
\subsection{The Goroff-Sagnotti operator ${\cal O}_3\equiv W_{\m\n\a\b}W^{\a\b\r\s}W_{\r\s}^{~~\m\n}$.}
The action for ${\cal O}_{GS}$ is
\be
S_{GS}\equiv \int d^4x \sqrt{|g|}W_{\m\n\a\b}W^{\a\b\r\s}W_{\r\s}^{~~\m\n}.
\ee 
It is instructive to work out the EM in arbitrary dimensions. They read
\begin{align}
	&V^{(1)}_{\m\n}=\frac{1}{2}g_{\m\n}\mathcal{O}_{GS}+\frac{6}{(n-2)}\Bigg[\frac{1}{n-1}\left(RW_{\m}^{~\a\b\l}W_{\n\a\b\l}-R_{\m\n}W^2\right)-R^{\a\b}W_{\m\a\l\t}W^{\l\t}_{~~\n\b}-\nonumber\\
	-&2\nabla^\t\nabla_\m (W_{\n}^{~\a\b\l}W_{\t\a\b\l})+g_{\m\n}\nabla^\r\nabla^\s( W_{\r}^{~\a\b\l}W_{\s\a\b\l})-(n-2)\nabla^\r\nabla^\s (W_{\m\r\a\b}W^{\a\b}_{~~\n\s})+\nonumber\\
	+&\left(\frac{(\nabla_\m\nabla_\n-g_{\m\n}\Box)}{n-1}W^2+\Box (W_{\m}^{~\a\b\l}W_{\n\a\b\l})\right)+R_\m^{~\t}W_{\n}^{~\a\b\l}W_{\t\a\b\l} \Bigg]-3W_{\m}^{~\a\b\l}W_{\n\a\r\s}W^{\r\s}_{~~\b\l}.\label{emw31}
\end{align}
The trace of \cref{emw31} corresponds to, 
\be\label{trace1} V^{(1)}=\frac{n-6}{2}W_{\m\n\a\b}W^{\a\b\r\s}W_{\r\s}^{~~\m\n}.\ee
We see that in any dimension different from the conformal one ($n=6$) a {\em necessary} condition for the EM to be satisfied is that
\be
{\cal O}_3:=W_{\m\n\a\b}W^{\a\b\r\s}W_{\r\s}^{~~\m\n}=0.
\ee

For the other contraction (which is independent of the former for $n\geq 6$), the story is quite similar

\be
S_{\tiny{Weyl^3}}\equiv \int d^4x \sqrt{-g}W^{\a~\b}_{~\g~\e} W^{\g~\e }_{~\r~\s}W^{\r~\s}_{~\a~\b}, \label{19} 
\ee
varying \cref{19} yields the EM,
\begin{align}
	&V^{(2)}_{\m\n}=-\frac{3}{n-2}\Bigg[\frac{1}{n-1}\left(2RW_{\m\a\b\l}W_\n^{~\b\a\l}+R_{\m\n}W_{\a\b\r\s}W^{\a\r\b\s}+(\nabla_\m\nabla_\n-g_{\m\n}\Box)(W_{\a\b\r\s}W^{\a\r\b\s}\right)-\nonumber\\
	&-2\nabla^\l\nabla_\m (W_{\n\a\b\t}W_\l^{~\b\a\t})+g_{\m\n}\nabla^\r\nabla^\s  (W_{\r\a\b\t}W_\s^{~\b\a\t})+\Box(W_{\m\a\b\t}W_\n^{~\b\a\t})-\nonumber\\
	&-2R^{\r\s}\left(W_{\m~\r}^{~\a~\b}W_{\n\b\s\a}-W_{\m~\n}^{~\a~\b}W_{\r\a\s\b}\right)\Bigg]+3R_{\m}^{~\a\l\b}W_{\n~\l}^{~\r~\s}W_{\a\r\b\s}-6W_{\m}^{~\a\l\b}W_{\n~\l}^{~\r~\s}W_{\a\r\b\s})+\nonumber\\
	&+\frac{1}{2}g_{\m\n}W^{\a~\b}_{~\g~\e} W^{\g~\e }_{~\r~\s}W^{\r~\s}_{~\a~\b}-3\nabla^\r\nabla^\s(W_{\m~\n}^{~\a~\b}W_{\r\b\s\a}) +3\nabla^\r\nabla^\s(W_{\m~\r}^{~\a~\b}W_{\n\b\s\a})	,
\end{align}
as in the previous case, the trace of the EM now corresponds to
\be
\label{trace2} V^{(2)}=\frac{n-6}{2}W^{\a~\b}_{~\g~\e} W^{\g~\e }_{~\r~\s}W^{\r~\s}_{~\a~\b},
\ee
with a similar necessary condition for the trace. 
We now see how this generalizes for $\mathcal{O}_{p\geq3}$.

\section{Solving the trace equation.}
Let us consider the action
\be
S_p\equiv \int d(vol) {\cal O}_p\equiv \int d^n x {\cal B}_p,
\ee
in arbitrary dimension $n$. Notice that we have defined ${\cal B}_p\equiv \sqrt{|g|}\, {\cal O}_p$
\par
Consider a linearized Weyl rescaling
\be
\d_W g_{\m\n}=\omega(x) g_{\m\n}.
\ee
Then
\be
\d_W S_p=\int d^n x \d_W {\cal B}_p=\int d^n x (n-2p)\omega {\cal B}_P=\int d^n x{\d S_p\over \d g_{\m\n}} \omega g^{\m\n}.
\ee
It follows that
\be
g^{\m\n}{\d S_p\over \d g_{\m\n}} =(n-2p){\cal B}_P, \label{25} 
\ee
which conveys the fact that the action is conformal\footnote{We stress that the utility of the trace equation lies on its simplicity which is directly tied to conformal invariance.}  in dimension $n=2p$.

This is the motivation for studying in some detail the trace equation  \eqref{25}, which will not always imply the vanishing of Weyl's tensor.
\par
Let us represent, following \cite{Matte}\footnote{
In fact, Matte's work precedes Petrov's  by one year, although he did not achieve a full classification of Weyl's tensor.
} and \cite{Pirani, Wheeler},
the 10 independent components of Weyl's tensor as a six-dimensional block matrix
\be
W^{\m\n}\,_{\r\s}=\begin{pmatrix}E&-B\\B&E\end{pmatrix},
\ee
where both $B$ and $E$ are three-dimensional, traceless, symmetric and real matrices.
Clearly\footnote{Here N takes care of index undercounting and its exact value is unmaterial for our purposes.},
\be
W^2=N\begin{pmatrix} E^2-B^2&-\left(EB+BE\right)\\BE+EB&E^2-B^2
\end{pmatrix}\quad\rightarrow \quad \text{tr} W^2=N \text{tr}\left(E^2-B^2\right).
\ee
Similarly,
\be
W^3=N \begin{pmatrix} E^3-EB^2-B^2 E-BEB&B^3-BE^2-E^2 B-BE^2-EBE\\BE^2-B^3+EBE+E^2 B&E^3-EB^2-B^2 E-BEB
\end{pmatrix},
\ee
where now, using the cyclic property of the trace,
\be
{\cal O}_{GS}=\text{tr} W^3=N\text{tr}\left(E^3-3 E B^2\right).
\ee
Given the fact that Schwarzschild spacetime, (as in fact all static, spherically symmetric spacetimes) have purely electric Weyl tensors, it can be easily shown that for them,
\be
{\cal O}_{GS}=0\Longrightarrow W_{\m\n\r\s}=0.
\ee
For purely magnetic spacetimes,  ${\cal O}_{GS}=0$ is given. Unfortunately, there are not many known purely magnetic exact solutions. There is a theorem stated in \cite{Mac} claiming that there are no purely magnetic Type D vacuum solutions of Einstein's equations.
\par
Starting with $\mathcal{O}_4$, the cyclic property of the trace is not enough to get a simple result.
To be more specific, we need some explicit results \cite{Pirani, Stephani} on Petrov's classification of spacetimes. They are frequently expressed in terms of the properties of the matrix
 \be
 Q\equiv E+i B.\label{Q} 
 \ee
The algebraically general spacetime is dubbed Petrov Type I. For these spacetimes, the general criterion is
 \be
 (Q-\l_1)(Q -\l_2)(Q-\l_3)=0,\label{32} 
 \ee
 (where $\sum \l_i=0$; and $\l_i\equiv e_i+i b_i$).
 The electric and magnetic Weyl tensors obey
 \be
 E=\begin{pmatrix}e_1&0&0\\0&e_2&0\\0&0&e_3\end{pmatrix},\quad\quad\mbox{and }\quad\quad B=\begin{pmatrix}b_1&0&0\\0&b_2&0\\0&0&b_3\end{pmatrix} \label{matrix} 
 \ee
 with $\sum e_i=\sum b_i=0$. It is plain that in this case, $[E,B]=0$ so the trace is given by
\be
\text{tr}\,W^3=2\,N\,\sum_i e_i(e_i^2-3 b_i^2).
\ee
Type D is a subtype that corresponds to $\l_1=\l_2=-2 \l_3$ in \cref{32}.
This particular Petrov Type is important for us because all spherically symmetric solutions \cite{Stephani} are either Type D or else Type O, (which means that the spacetime is conformally flat).
\par
The general result for the all-important Types I and D is obtained from the above discussion;
 \be
 \text{tr}\, W^{2m}=2 \sum_{k=0}^m (-1)^k \binom{2m}{2k}\text{tr}\,\left(E^{2(m-k)} B^{2k}\right),
 \ee
 \be
 \text{tr}\, W^{2m+1}=2 \sum_{k=0}^m (-1)^k \binom{2m+1}{2k}\text{tr}\,\left(E^{2(m-k)+1} B^{2k}\right).
 \ee
This gives immediately the conditions for the vanishing of the trace in terms of products of the electric and magnetic eigenvalues $\{e_i,\,b_i\}$.

\section{Spherically symmetric spacetimes (S3).}

Let us now particularize to the case of a spherically symmetric spacetime S3 with metric tensor, 
\begin{equation}
ds^{2} =B(t,r)dt^{2} -A(t,r)dr^{2} -r^{2}\,d\Omega_{2}^{2} .
\end{equation}
In \cite{Alvarez2022} we were mostly concerned with $ \text{tr}\, W^{3}  $ but we also made the observation that for this particular S3 spacetimes, 
\begin{equation}
\mathcal{O}_4 := \tensor{W}{^{\alpha} ^{\beta} _\gamma_{\delta} } \tensor{W}{^{\gamma} ^{\delta} _\kappa _{\zeta} }  \tensor{W}{^{\kappa} ^{\zeta}  _\lambda_\sigma  }\tensor{W}{^{\lambda} ^{\sigma} _{\alpha} _{\beta} } ,
\end{equation}
is proportional to the lagrangian of Conformal Gravity (CG) in \cref{CG} . 

As a matter of fact, \cite{Alvarez2022,Deser} Weyl's tensor for an S3 depends on a unique combination of $  A(t,r),\,B(t,r) $ and their derivatives. Using the notation of \cref{sec:A1}, it so happens that
\begin{equation}
	\sqrt{-g}\mathcal{O}_2={\sqrt{r^4 A B \sin ^2(\theta )}\over 12 r^4 A^4 B^4}\,G^2,
\end{equation}
and,
\begin{equation}
	\sqrt{-g}\mathcal{O}_4={\sqrt{r^4 A B \sin ^2(\theta)} \over 576 r^8 A^8 B^8}\,G^4.
\end{equation}
However, we stressed that CG was different from any other $  S_p  $ action in which it contained solutions not present for the $  p\neq 2 $ cases. 
It is important to understand how the $  p=4  $ fails to capture such solutions even when this similarity is present. 

This simple relationship between both lagrangians does not carry over to the corresponding solutions. As shown in \cite{Alvarez2022}, the solutions for CG contain those of $S_4$  but not conversely. 
Of course, for CG the trace equation \eqref{12}  does not yield any restriction on the conformal case $n=4$. On the other hand, for the EM for $S_4$, the trace condition applies. Therefore,  the only way in which a metric can be a solution to both CG and  quartic gravity is to 
 satisfy, in addition 
 \begin{equation}
 W^{2}=\tensor{W}{^\alpha^\beta_\gamma_\delta}\tensor{W}{^\gamma^\delta_\alpha_\beta}=0 \Rightarrow G=0\Rightarrow \tensor{W}{^\alpha_\beta_\gamma_\delta}=0, \label{40} 
 \end{equation}
 i.e., it has to be conformally-flat.
In fact for any $p\neq2$, S3 solutions are determined by \cref{40}.
\section{Conclusions.}
Let us finish with a couple of examples that depart from spherical symmetry and thus from the theorem proven in \cite{Alvarez2022}. \\
\begin{itemize}
	\item First we consider the spacetime of p-p waves with line element \cite{Stephani},
	\begin{equation}
		ds^2=-2 \text{du}^2 H(x,y,u)+2 \text{du} \text{dv}-\left(\text{dx}^2+\text{dy}^2\right).
	\end{equation}
	To solve Einstein's equations in vacuum, $H(x,y,u)$ has to satisfy
	\begin{equation}
		\tensor{R}{_\mu_\nu} =0\quad\Leftrightarrow\quad\partial_x^2\;H(x,y,u)+\partial_y^2\;H(x,y,u)=0.\label{empp} 
	\end{equation}
The above-introduced analysis yields electric and magnetic components of the Weyl tensor corresponding to, 
\begin{equation}
	E=\frac{1}{2 H(x,y,u)+2}\left(
		\begin{array}{ccc}
		 0 & 0 & 0 \\
		 0 & \partial_x^2\;H(x,y,u) & \partial_x\partial_y\;H(x,y,u)\\
		 0 & \partial_x\partial_y\;H(x,y,u) & -\partial_x^2\;H(x,y,u) \\
		\end{array}
		\right),\label{Epp} 
\end{equation}
and, 
\begin{equation}
	B=\frac{1}{2 H(x,y,u)+2}\left(
		\begin{array}{ccc}
		 0 & 0 & 0 \\
		 0 & -\partial_x\partial_y\;H(x,y,u) & \partial_x^2\;H(x,y,u)\\
		 0 & \partial_x^2\;H(x,y,u) & \partial_x\partial_y\;H(x,y,u) \\
		\end{array}
		\right).\label{Mpp} 
\end{equation}
In \cref{Mpp,Epp}, \cref{empp} has been used. 
Therefore, we see that p-p waves are everywhere of Petrov Type N. 
This implies,
\begin{equation}
	Q^2=0,
\end{equation}
which translates into the trace condition being satisfied for them, i.e.,
\begin{equation}
	\mbox{tr}\,W^3\;\propto\;  \mbox{tr}\;Q^3=0.
\end{equation}
Thus, p-p waves comprise an example of a spacetime satisfying the necessary condition to be a solution to the EM. 
Here, direct computation allows us to see that this spacetime also satisfies the Goroff-Sagnotti EM, \cref{emw31}.

In \cite{Alvarez2022} it was proven that no spherically symmetric spacetime can simultaneously solve Einstein's and Goroff-Sagnotti's EM. It is interesting to remark that p-p waves are the only example of a solution to both EM we have been able to identify to this date.
\item As a second non-spherically symmetric nor static example, we consider Kerr-Newmann spacetime, whose line element, in Boyer-Lindquist coordinates \cite{Oneill}  reads, 
\begin{equation}
	ds^2 = \frac{\Delta}{\rho^2}\left(dt - a \sin^2\theta \,d\phi \right)^2 -\frac{\sin^2\theta}{\rho^2}\Big[\left(r^2+a^2\right)\,d\phi - a \,dt\Big]^2 - \frac{\rho^2}{\Delta}dr^2 - \rho^2 \,d\theta^2,\label{Kerr} \\
\end{equation}
where,
\begin{equation}
	\Delta:={r}^2-r r_s+a^2+r_q^2,\qquad\qquad\mbox{and }\qquad\qquad\rho:=a^2 \cos ^2(\theta )+r^2.
\end{equation}
As usual, $a$ is related to the angular momentum, $r_s$ is the Schwarzschild radius, related to the mass and $r_q$ is related to the electric charge \footnote{	We set $r_q=0$ such that the line element in \cref{Kerr} is a solution to the vacuum Einstein's EM. } of the solution. 

In this case, the electric and magnetic matrices are, 
\begin{align} &E=e(r,\theta)\begin{pmatrix}
	1&0&0\\
	0&-\frac{1}{2}&0\\
	0&0&-\frac{1}{2}
\end{pmatrix},\,\mbox{with }\,e(r,\theta)=\frac{4r_s r \bigl(3 a^2 - 2 r^2 + 3 a^2 \cos(2 \theta)\bigr)}{\bigl(a^2 + 2 r^2 + a^2 \cos(2 \theta)\bigr)^3},\label{eker} \\
&B=b(r,\theta)\begin{pmatrix}
	1&0&0\\
	0&-\frac{1}{2}&0\\
	0&0&-\frac{1}{2}
\end{pmatrix},\,\mbox{with }b(r,\theta)=\frac{4r_s a \cos(\theta) \bigl(a^2 - 6 r^2 + a^2 \cos(2 \theta)\bigr)}{\bigl(a^2 + 2 r^2 + a^2 \cos(2 \theta)\bigr)^3}.\label{bker} 
\end{align}
This corresponds to a Petrov Type D spacetime, following the above discussion,  
\begin{equation}
	\text{tr} W^{3}=\frac{3}{4}e(r,\theta)(e(r,\theta)^2-3b(r,\theta)^2)=0.\label{trkerr} 	
\end{equation}
The trace in \cref{trkerr} is nonvanishing\footnote{Unless $r_s=0$.} for arbitrary $r$ values.  
We conclude that an uncharged Kerr spacetime with mass cannot solve the EM for the Goroff-Sagnotti counterterm. This we believe to be an important result. Not only Schwarzschild but also its most important stationary extension, namely Kerr (widely believed to be the general endpoint of gravitational collapse), is incompatible with the Goroff-Sagnotti lagrangian deformation.  
\end{itemize}
Let us finish with a quick summary of this work. In this letter, we have studied the effects of including generalizations of the Goroff-Sagnotti operator on Einstein-Hilbert's action for GR. 
Specifically, we have focused on the equations of motion and obtained one necessary condition for their solutions. 
\par
In detail, the  operators we have studied are built of $p$ contractions of the Weyl tensor, symbolically, 
\begin{equation}
\int d^nx\;\sqrt{|g|}\mathcal{O}_p:=\int d^nx\;\sqrt{|g|} \text{tr}\, W^p.
\end{equation}
These are conformally invariant in dimension $n=2p$. 
Using the so-called trace equation and the Petrov classification of Weyl's tensor, we have found a necessary condition for the solutions of the EM in the general case.
\par
In the more simple instance of spherically symmetric spacetimes, which motivated this study originally \cite{Alvarez2022},
we are able to find the general solution of the EM. This is because, in this case, the trace condition becomes a necessary and sufficient condition for a spacetime to solve for the complete EM.
\par
Examples for the \textit{less trivial} p-p waves and Kerr-spacetime are explicitly worked out. p-p waves are found to be a solution to the EM of GR with the $p=3$ operator included.

\section{Acknowledgements.}
We have enjoyed fruitful discussions with Prof. Senovilla. We acknowledge partial financial support by the Spanish MINECO through the Centro de Excelencia Severo Ochoa Program  under Grant CEX2020-001007-S  funded by MCIN/AEI/10.13039/501100011033
All authors acknowledge the European Union's Horizon 2020 research and innovation programme under the Marie Sklodowska-Curie grant agreement No 860881-HIDDeN and also by Grant PID2019-108892RB-I00 funded by MCIN/AEI/ 10.13039/501100011033 and by ``ERDF A way of making Europe''.
This project has received funds/support from the European Union's Horizon Europe programme uncer Marie Sklodowska-Curie Actions-Staff Exchanges (SE) grant agreement No 101086085-ASYMMETRY.
\appendix
\section{The S3 Weyl tensor.} \label{sec:A1} 
Direct computation of the  Weyl tensor for S3 spacetimes yields
\be
 W^{\m\n}_{~~\r\s}=\frac{G}{12A^2B^2}\,  C^{\m\n}_{~~\r\s},\tag{A.1}
\ee
In particular,
\begin{align}
    G(A,A',B,B',\cdots)&:=A \left(r^2 \left(B'^2-\dot{A} \dot{B}\right)+2 r B \left(r \ddot{A}+B'-r B''\right)-4 B^2\right)+\nonumber\\
    &+r B \left(A' \left(r B'-2 B\right)-r \dot{A}^2\right)+4 A^2 B^2, \tag{A.2}
 \end{align}
where  $\dot{f}=\partial_t f(r,t)$ and $f'=\partial_r f(r,t)$.
$C^{\m\n}_{~~\r\s}$ is a tensor independent of the metric. 
In detail;
\begin{align}
C^{01}_{~~\a\b}=-C^{10}_{~~\a\b}&=\begin{pmatrix}
	0&-1&0&0\\
	1&0&0&0\\
	0&0&0&0\\
	0&0&0&0\end{pmatrix},\qquad
&&C^{02}_{~~\a\b}=-C^{20}_{~~\a\b}=\begin{pmatrix}
	0&0&\frac{1}{2}&0\\
	0&0&0&0\\
	-\frac{1}{2}&0&0&0\\
	0&0&0&0\end{pmatrix},\nonumber\\
C^{03}_{~~\a\b}=-C^{30}_{~~\a\b}&=\begin{pmatrix}
	0&0&0&\frac{1}{2}\\
	0&0&0&0\\
	0&0&0&0\\
	-\frac{1}{2}&0&0&0\end{pmatrix},\qquad
	&&C^{12}_{~~\a\b}=-C^{21}_{~~\a\b}=\begin{pmatrix}
	0&0&0&0\\
	0&0&\frac{1}{2}&0\\
	0&-\frac{1}{2}&0&0\\
	0&0&0&0\end{pmatrix},\nonumber\\
C^{13}_{~~\a\b}=-C^{31}_{~~\a\b}&=\begin{pmatrix}
	0&0&0&0\\
	0&0&0&\frac{1}{2}\\
	0&0&0&0\\
	0&-\frac{1}{2}&0&0\end{pmatrix},\qquad
&&C^{23}_{~~\a\b}=-C^{32}_{~~\a\b}=\begin{pmatrix}
	0&0&0&0\\
	0&0&0&0\\
	0&0&0&-1\\
	0&0&1&0\end{pmatrix}.\tag{A.3}
\end{align}
This Weyl tensor is purely electric, with 

\be E=\frac{G}{12A^2B^2}\begin{pmatrix}
	-1& 0&0\\
	0 & \frac{1}{2}&0\\
	0&0&\frac{1}{2}
\end{pmatrix}.\tag{A.4}\ee

	
\end{document}